\documentclass{optica-article}

\journal{opticajournal} 

\articletype{Research Article}

\usepackage{tikz}
\usetikzlibrary{intersections}
\usepackage[colorinlistoftodos]{todonotes}
\usepackage{subcaption}  
\usepackage{mwe}         
\usetikzlibrary{decorations.pathreplacing}
\usetikzlibrary{arrows.meta, positioning, shapes.geometric}

\begin{document}

\title{Fundamentals of Optical Fiber Sensing Schemes Based on Coherent Optical Time Domain Reflectometry: Signal Under Dynamic Temperature Conditions}

\author{
Roman~Ermakov,\authormark{1,*} 
Huwei~Wang,\authormark{1}
Florian~Azendorf, \authormark{2}
André~Sandmann,\authormark{2}
Juan~M.~Marin,\authormark{1} 
Francesco~Da~Ros,\authormark{1} 
and 
Darko~Zibar\authormark{1}
}

\address{
\authormark{1}DTU Electro, Technical University of Denmark (DTU), DK-2800 Kgs. Lyngby, Denmark\\
\authormark{2}Adtran Networks SE, Märzenquelle 1-3, 98617 Meiningen, Germany
}

\email{\authormark{*}romane@dtu.dk} 


\begin{abstract*} 
We present a theoretical, algorithmic, and experimental study of temperature sensing using \(\phi\)-OTDR with coherent detection. A physics-based model is developed to relate the measured Rayleigh backscattered signal to temperature variations along the fiber, showing that the phase evolution encodes the cumulative temperature change between the interrogator and the sensing location, while the amplitude exhibits only local sensitivity. Based on this insight, we propose robust algorithms for temperature-event detection and temperature-profile reconstruction. Experimental results demonstrate reliable recovery of temperature-induced perturbations in standard single-mode fibers using coherently detected \(\phi\)-OTDR.
\end{abstract*}

\section{Introduction}
Phase-sensitive optical time-domain reflectometry ($\phi$-OTDR) has emerged as a state-of-the-art technique that uses deployed optical fibers as distributed sensors~\cite{ref1}. By exploiting the sensitivity of the Rayleigh backscattered signal to external perturbations, $\phi$-OTDR enables distributed monitoring of temperature~\cite{ref2}, strain~\cite{ref3}, and vibrations~\cite{ref4} in large-scale infrastructures such as overhead power transmission lines~\cite{ref5} and telecommunication networks~\cite{ref6}.

This sensitivity arises from the Rayleigh backscattering phenomenon, in which a coherent light pulse propagating through an optical fiber is scattered by microscopic fluctuations of the fiber’s refractive index~\cite{ref8}. These scattering centers form interference patterns, known as Rayleigh speckles~\cite{ref9,ref10}, whose amplitude and phase change in response to local variations in temperature and strain. For example, a temperature change in a section of the fiber under test (FUT) induces both a local refractive index variation~\cite{ref11} and thermal expansion of the fiber~\cite{ref12}, thereby modifying the optical path length and, consequently, the phase of the backscattered field. The associated change in the relative positions of the scattering centers leads to a redistribution of the backscattered amplitude.

For this reason, $\phi$-OTDR systems that directly measure the optical phase of the backscattered signal—rather than only its intensity—provide more reliable information about external perturbations. Among the various phase-resolved configurations~\cite{ref13,ref14,ref15}, the coherent-receiver-based $\phi$-OTDR scheme~\cite{ref16,ref17} is gaining traction in practical applications~\cite{ref1,ref4,ref5}. This approach offers a favorable trade-off between sensing range, spatial resolution, and sensing bandwidth, while significantly improving the signal-to-noise ratio (SNR) through the use of pulse-coding techniques~\cite{ref18, ref19}. It enables extended sensing ranges~\cite{ref20,ref3,ref1,Waagaard:21} with sufficient SNR, supports reuse of existing telecommunication single-mode fibers as the sensing medium~\cite{ref21}, and allows cost-effective system implementation.

However, a fundamental gap remains in the current body of work: for coherent $\phi$-OTDR there is no general, physics-based mechanism that quantitatively relates changes in the Rayleigh backscattered signal to external temperature variations. Existing approaches to temperature sensing in this configuration are predominantly empirical. The method in~\cite{ref22}, based on inverse low-pass filtering of the phase rate, relies on specific calibration procedures and is tightly coupled to the particular experimental scenario. Another study~\cite{ref23} reports good agreement between measured phase differences and calculated temperature changes, but the results are obtained in a highly specialized setup and rely on an oversimplified model that neglects the temperature dependence of the refractive index. Other distributed temperature sensing schemes, such as wavelength-swept~\cite{ref24,ref25} and chirped-pulse~\cite{ref26} $\phi$-OTDR, or hybrid $\phi$-OTDR/Raman OTDR systems~\cite{ref27}, can provide quantitative temperature information, but they rely on different system architectures and typically do so at the expense of measurement speed, long-range SNR, or overall system cost and complexity.

In this work, we develop a theoretical framework describing how the Rayleigh backscattered signal in $\phi$-OTDR systems based on coherent detection is modified by external perturbations along the sensing fiber. Building on the static Rayleigh model of~\cite{main_model}, we derive an analytical description that maps changes in physically relevant fiber parameters to measurable quantities of the backscattered signal, enabling quantitative interpretation and accurate modeling of external influences. Specializing this framework to temperature, we obtain a closed-form expression for the Rayleigh signal variations induced by temperature changes and, by isolating the phase contribution, derive a practical relation between the measured optical phase and the local temperature change. This phase–temperature relation forms the basis for two algorithms: one for detecting and localizing temperature events, and another for reconstructing the temporal evolution of the temperature change along the fiber. Experimentally, we validate these algorithms on standard single-mode fiber, achieving temperature rate uncertainties on the order of $0.05^{\circ}\mathrm{C}/\mathrm{s}$ without complex calibration. It is worth mentioning that this work provides a comprehensive generalization of our previously published temperature-recovery algorithm in~\cite{Ermakov}.

This paper is organized as follows. In Section~\ref{sec:theory}, Subsection~\ref{sec:coherent_detection}, we introduce the measurable quantities used in the analysis and experiments. In Subsection~\ref{sec:static_model}, we then briefly review the theoretical model of the Rayleigh backscattered signal in a stationary FUT~\cite{main_model}, expressed in terms of intrinsic fiber parameters. Next, in Subsection~\ref{sec:dynamic_model}, we extend this model by incorporating the temperature dependence of these parameters and analyze how temperature variations manifest in the backscattered signal. In Section~\ref{sec:experiment}, we describe the experimental setup and procedures used to verify the model. Finally, in Section~\ref{sec:results}, we present a temperature reconstruction algorithm based on the derived expressions, enabling quantitative recovery of both the temperature rate along the FUT and the temperature change in the time domain.  

\section{Theory}\label{sec:theory}
\subsection{Digital Coherent Detection of the Backscattered Signal}\label{sec:coherent_detection}

\begin{figure}[t]
    \centering\includegraphics[width=14cm]{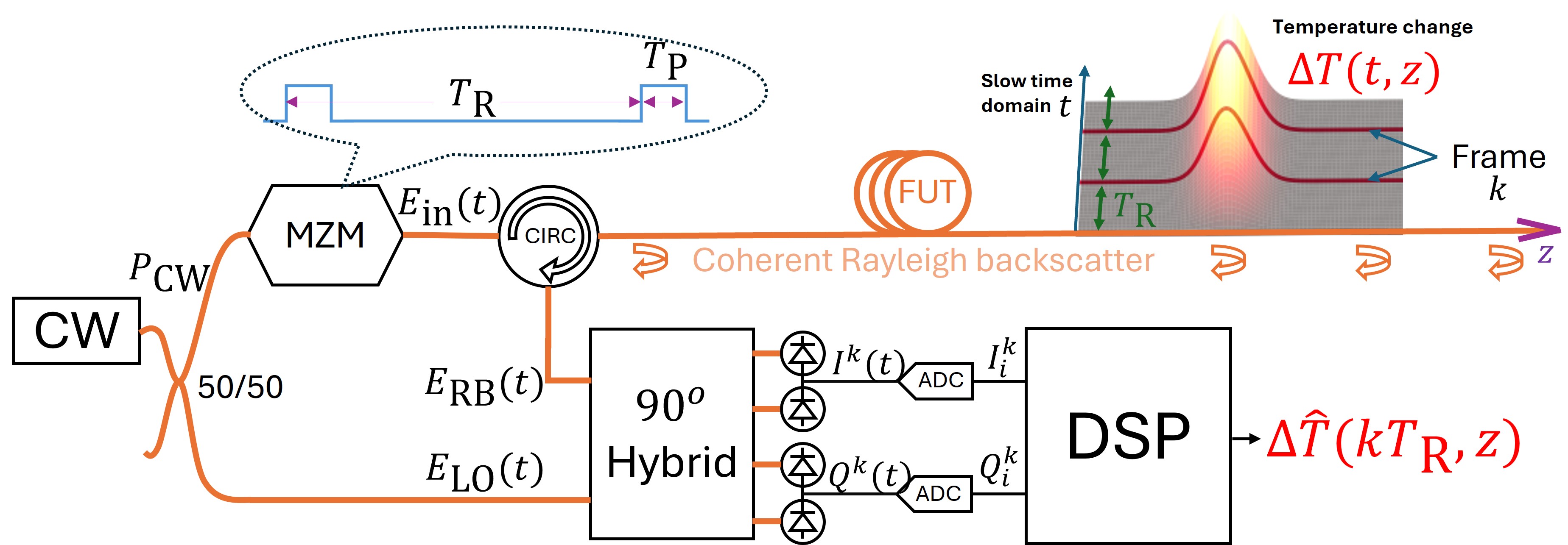}
    \caption{$\phi$-OTDR measurement setup: CW: a continuous-wave laser source; MZM: electro-optic modulator driven by pulse pattern generator; CIRC: circulator; FUT: fiber under test; $90^{\circ}$ Hybrid: coherent optical receiver; balanced photodetectors; ADC: analog-to-digital converter; and DSP block.}
    \label{fig:figure_1}
\end{figure}

The main objective of this section is to establish a theoretical framework and digital signal processing (DSP) pipeline for modeling and estimating a locally induced, time-varying temperature perturbation \(\Delta T(t,z) = T(t,z) - T_{\mathrm{init}}(z)\), where \(T(t,z)\) denotes the temperature along the FUT at position \(z\) and time \(t\), and \(T_{\mathrm{init}}(z)\) is the initial temperature distribution at the start of the measurement. The framework is based on a standard \(\phi\)-OTDR configuration, illustrated in Fig.~\ref{fig:figure_1}. The model assumes single-polarization propagation that remains constant along the sensing fiber. Its primary outcome is the estimation of the induced temperature-change snapshot along the FUT at a specific time instant \(t_0\), denoted by \(\Delta \hat{T}(t_0,z)\).

The experimental setup, shown in Fig.~\ref{fig:figure_1}, employs a continuous-wave (CW) laser source operating at angular frequency \(\omega\) with optical power \(P_{\mathrm{CW}}\). The laser is assumed to be free of phase noise and relative intensity noise. The output is split by a \(50/50\) coupler, and one arm is fed into a Mach--Zehnder modulator (MZM) driven by a pulse pattern generator (PPG). This modulation produces a train of rectangular optical pulses with pulse width \(T_{\mathrm{P}}\) and repetition period \(T_{\mathrm{R}}\), corresponding to a repetition rate \(F_{\mathrm{R}} = 1/T_{\mathrm{R}}\).

The pulse repetition period must be longer than the fiber round-trip time \(\tau_{\mathrm{f}} = \frac{2 n_g L_{\mathrm{FUT}}}{c}\), where \(L_{\mathrm{FUT}}\) is the length of the sensing fiber and \(n_g\) is the group refractive index. The modulated optical signal is routed through an optical circulator and launched into the sensing fiber. The resulting input optical field can be expressed as
\begin{equation}
E_{\mathrm{in}}(t) =
\sqrt{P_{\mathrm{in}}}\,
\mathrm{e}^{\mathrm{j}(\omega t + \phi_0)}
\sum_{k=0}^{\infty}
\operatorname{rect}\!\left(\frac{t - k T_{\mathrm{R}}}{T_{\mathrm{P}}}\right),
\end{equation}
where \(\phi_0\) is the initial optical phase and \(P_{\mathrm{in}} = 0.5\,P_{\mathrm{CW}}\) denotes the peak optical power of the launched pulses, with the factor \(0.5\) accounting for the \(50/50\) power splitting. Under the assumption of ideal pulse carving, the average optical power of the launched pulse train is \(P_{\mathrm{avg}} = P_{\mathrm{in}}\frac{T_{\mathrm{P}}}{T_{\mathrm{R}}}
= 0.5\,\frac{T_{\mathrm{P}}}{T_{\mathrm{R}}}\,P_{\mathrm{CW}}\).

Due to Rayleigh backscattering in the optical fiber, the launched signal \(E_{\mathrm{in}}(t)\) generates a backscattered optical field \(E_{\mathrm{RB}}(t)\), which can be written as~\cite{journal:Bao2021Rayleigh}
\begin{equation}
\begin{aligned}
E_{\mathrm{RB}}(t) =
\sqrt{P_{\mathrm{RB}}}\,
\mathrm{e}^{\mathrm{j}(\omega t + \phi_0)}
\exp\!\left\{-\alpha z(t)\right\}
\sum_{k=0}^{\infty}
A^{k}(t - k T_{\mathrm{R}})\,
\exp\!\left\{\mathrm{j}\,\phi^{k}(t - k T_{\mathrm{R}})\right\},
\end{aligned}
\end{equation}
where \(A^{k}(t)\) and \(\phi^{k}(t)\) denote the time-varying amplitude and phase of the backscattered signal associated with the \(k\)-th probe pulse, \(\alpha\) is the fiber attenuation coefficient, and \(z(t)\) is the spatial coordinate from which the signal originated at time \(t\). To relate time \(t\) to the spatial coordinate \(z(t)\), we follow the approach in~\cite{main_model}:
\begin{equation}
\label{eq:z(t)}
t = \frac{2}{c} \int_0^{z(t)} n_g(x)\,\mathrm{d}x,
\end{equation}
which defines the spatial coordinate \(z(t)\) along the sensing fiber through the local group index \(n_g(x)\).

The average Rayleigh backscattered power \(P_{\mathrm{RB}}\) is proportional to the launched optical power and can be expressed as~\cite{journal:Bao2021Rayleigh}
\[
P_{\mathrm{RB}} = 0.5\,v_g T_{\mathrm{P}}\,\alpha_{\mathrm{R}} B\,P_{\mathrm{in}},
\]
where the factor \(0.5\,v_g T_{\mathrm{P}} = \frac{c T_{\mathrm{P}}}{2 n_g}\) represents the effective spatial length of the scattering zone defined by the probe pulse duration, i.e., the fiber segment contributing to the measured Rayleigh backscattered signal. Here, \(\alpha_{\mathrm{R}}\) denotes the distributed Rayleigh scattering coefficient (in \(\mathrm{m}^{-1}\)), typically on the order of \(0.15~\mathrm{dB/km}\) at a wavelength of \(1.55~\mu\mathrm{m}\)~\cite{ref30}, and \(B\) is the capture fraction accounting for the portion of scattered light coupled back into the guided mode~\cite{ref31}.

The second arm of the \(50/50\) coupler is used as a local oscillator (LO), described by
\begin{equation}
\label{eq:E_LO}
E_{\mathrm{LO}}(t) =
\sqrt{P_{\mathrm{LO}}}\,
\mathrm{e}^{\mathrm{j}(\omega t + \phi_{\mathrm{LO}})},
\end{equation}
where \(P_{\mathrm{LO}} = 0.5\,P_{\mathrm{CW}}\) and \(\phi_{\mathrm{LO}}\) is a constant phase offset. The backscattered signal \(E_{\mathrm{RB}}(t)\) and the LO field \(E_{\mathrm{LO}}(t)\) are combined in a single-polarization \(90^{\circ}\) optical hybrid and detected using self-homodyne coherent detection with balanced photodetectors. The resulting complex baseband signal can be written as
\begin{equation}
\label{eq:I+Q}
I(t) + \mathrm{j} Q(t) =
\eta \sqrt{P_{\mathrm{RB}} P_{\mathrm{LO}}}\,
\mathrm{e}^{\mathrm{j}(\phi_{\mathrm{RB}} - \phi_{\mathrm{LO}})}
\exp\!\left\{-\alpha z(t)\right\}
\sum_{k=0}^{\infty}
\bar{A}^{k}(t - k T_{\mathrm{R}})\,
\exp\!\left\{\mathrm{j}\,\bar{\phi}^{k}(t - k T_{\mathrm{R}})\right\},
\end{equation}
where \(\eta\) denotes the responsivity of the (assumed identical) photodetectors. It is important to note that, due to the finite bandwidth of the detection chain, the signal \(I(t) + \mathrm{j} Q(t)\) at time \(t\) corresponds to a low-pass-filtered version of the optical interference term rather than an instantaneous measurement of the optical field. To reflect this effect, the amplitude and phase terms in Eq.~\eqref{eq:I+Q} are written as \(\bar{A}^{k}(\cdot)\) and \(\bar{\phi}^{k}(\cdot)\), emphasizing that they represent bandwidth-limited (electrically filtered) quantities. Special attention is required for the phase difference \(\phi_{\mathrm{RB}} - \phi_{\mathrm{LO}}\), which under our assumptions is taken to be independent of time \(t\). However, due to the infinite sum over the pulse index \(k\), laser phase noise cannot be neglected over macroscopic time scales. Therefore, throughout Section~\ref{sec:theory} we assume this phase difference to be zero, and we return to its impact when developing the reconstruction algorithm in Section~\ref{sec:results}.

The received signal in Eq.~\eqref{eq:I+Q} is digitized by a front-end analog-to-digital converter (ADC) at sampling instants \(t_i = i T_{\mathrm{S}}\), where \(T_{\mathrm{S}}\) is the sampling interval and \(F_{\mathrm{S}} = 1/T_{\mathrm{S}}\) is the sampling rate. Because the contributions associated with successive probe pulses in Eq.~\eqref{eq:I+Q} are separated by the repetition period \(T_{\mathrm{R}}\) (with repetition rate \(F_{\mathrm{R}} = 1/T_{\mathrm{R}}\)) and the usual operating condition \(T_{\mathrm{R}} \gg T_{\mathrm{S}}\) holds, it is convenient to represent the sampled signal using two indices. The index \(i\) corresponds to the \emph{fast-time} scale, describing the temporal evolution of the received signal within a single probe pulse, while the index \(k\) corresponds to the \emph{slow-time} scale, describing the evolution of the signal across successive probe pulses. In this representation, \(i = 0, 1, 2, \ldots, i_{\mathrm{max}}\), where \(i_{\mathrm{max}} = T_{\mathrm{R}}/T_{\mathrm{S}} = F_{\mathrm{S}}/F_{\mathrm{R}}\).

\begin{figure}[t!]
    \centering\includegraphics[width=12cm]{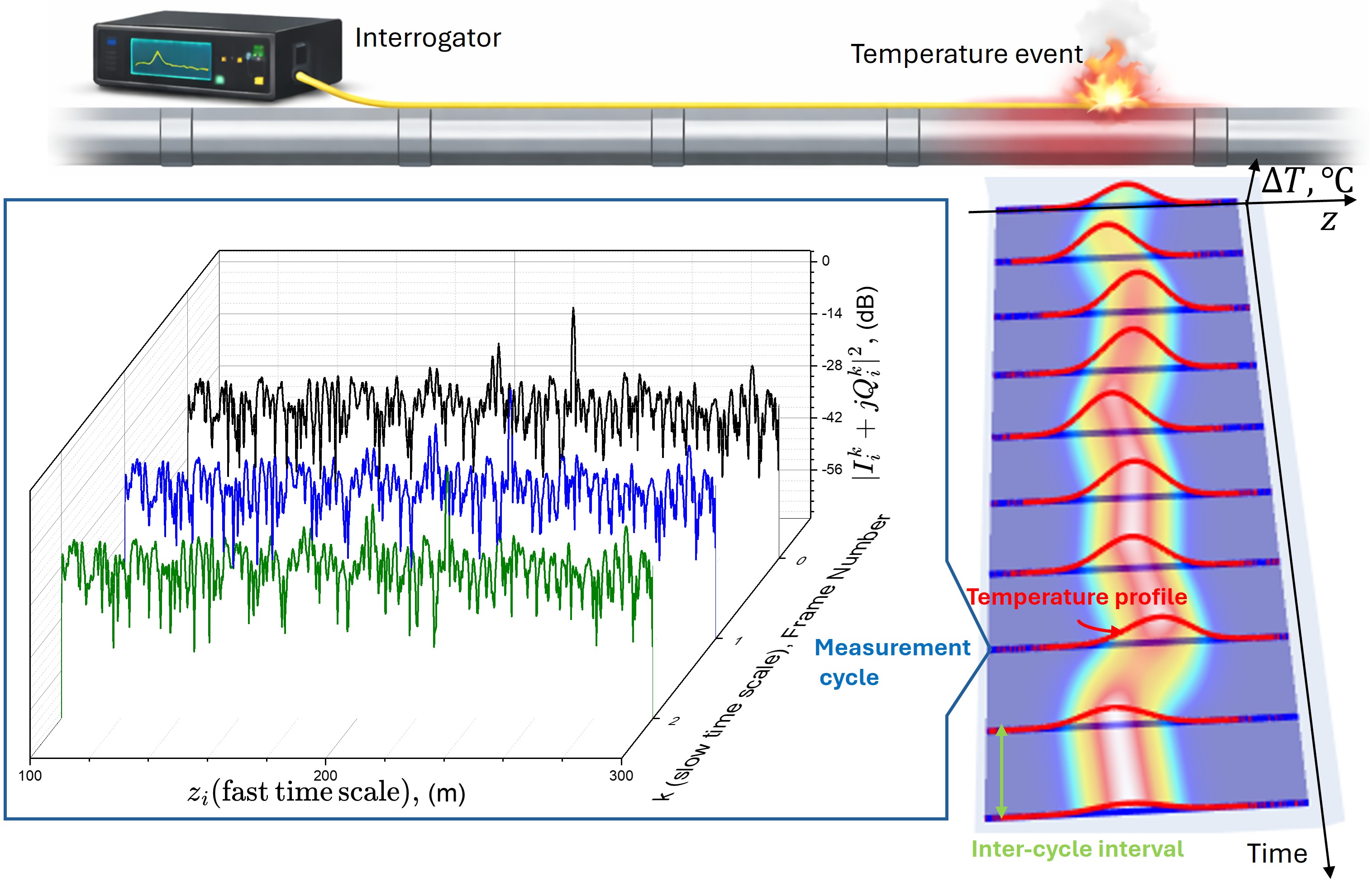}
    \caption{$\phi$-OTDR measurement process over macroscopic time.}
    \label{fig:measurements}
\end{figure}

Under this notation, the discrete-time version of Eq.~\eqref{eq:I+Q} for the \(k\)-th probe pulse (frame) can be expressed as
\begin{equation}
\label{eq:I+Q_digital}
I_i^{k} + \mathrm{j} Q_i^{k} =
\eta \sqrt{P_{\mathrm{RB}} P_{\mathrm{LO}}}\,
\bar{A}_i^{k}\,
\exp\!\left( \mathrm{j}\,\bar{\phi}_i^{k} \right)
\exp\!\left( -\alpha z_i \right),
\end{equation}
where \(i = 0,1,2,\ldots,i_{\mathrm{max}}\) is the fast-time sample index and the corresponding spatial coordinate \(z_i\) is obtained from the discrete version of Eq.~\eqref{eq:z(t)}:
\begin{equation}
\label{eq:delay2}
t_i = \frac{2}{c} \int_{0}^{z_i} n_g(x)\,\mathrm{d}x = i T_{\mathrm{S}},
\end{equation}
as given in Eq.~(9) of~\cite{main_model}. An example of a collection of frame fingerprints \(|I_i^{k} + \mathrm{j} Q_i^{k}|^2\) from Eq.~\eqref{eq:I+Q_digital} under local temperature perturbation is shown in Fig.~\ref{fig:measurements}.

Under the assumption of a uniform group refractive index along the sensing fiber, this relation simplifies to \(z_i = \frac{c}{2 n_g} i T_{\mathrm{S}}\), which shows that the fast-time index \(i\) corresponds to equidistant sampling points along the sensing fiber as shown at Fig.~\ref{fig:scatters_distribution}.

\subsection{Model of the Backscattered Signal Under Static Conditions}\label{sec:static_model}
Our theoretical framework builds on the model in \cite{main_model} and extends it to account for non-static behaviour of the sensing fiber. In this work, we restrict our attention to temperature-induced effects. To proceed, we briefly recall the main concepts necessary for the remainder of the paper.

The Rayleigh backscattered signal can be described as a superposition of electromagnetic waves scattered from individual inhomogeneities embedded in the fiber material. These scattering centers are randomly distributed along the fiber, fixed in position, and characterized by their locations \(z_{m}\) and scattering amplitudes \(a_{m}\), where \(m\) denotes the scatterer index and \(z_{m+1} > z_m\) for all \(m\). The amplitude \(a_{m}\) determines the fraction of light that is backscattered and is a positive random variable physically related to the scatterer size. Knowledge of the set \(\{z_m, a_m\}\) alone is not sufficient to predict the backscattered signal; the refractive index distribution \(n(x)\) along the fiber is also required. Thus, the state of the sensing fiber can be fully described by two components:
\begin{equation*}
    \left\{
        \begin{aligned}
            \{z_m, a_m\}, &\quad m \in \{1,2,\dots,M\}; \\
            n(z),
        \end{aligned}
    \right.
\end{equation*}
where \(M\) denotes the total number of scattering centers in the sensing fiber.

\begin{figure}[t!]
\centering
\begin{tikzpicture}[x=\linewidth, y=1cm]

  \pgfmathsetseed{2}

  \def\tickH{0.13}
  \def\N{120}

  \def\S{0.025}
  \def\L{0.20}
  \def\B{0.18}

  \pgfmathsetmacro{\a}{\S}
  \pgfmathsetmacro{\b}{\S+\L}
  \pgfmathsetmacro{\c}{\S+\L+\B}
  \pgfmathsetmacro{\d}{\S+2*\L+\B}
  \pgfmathsetmacro{\e}{\S+2*(\L+\B)}
  \pgfmathsetmacro{\f}{\S+2*(\L+\B)+\L}

  \pgfmathsetmacro{\mOne}{(\a+\b)/2}
  \pgfmathsetmacro{\mTwo}{(\c+\d)/2}
  \pgfmathsetmacro{\mThree}{(\e+\f)/2}

  \pgfmathsetmacro{\Dsample}{\mTwo-\mOne}
  \pgfmathsetmacro{\mZero}{\mOne-\Dsample}
  \pgfmathsetmacro{\mFour}{\mThree+\Dsample}

  \pgfmathsetmacro{\xm}{(\b+\c)/2}

  \def\yFiber{0}
  \def\yTime{-1.35}
  \def\yArrowTop{0.2}
  \def\yBrace{-0.42}

  \colorlet{blueStrong}{blue}
  \colorlet{blueFaint}{blue!40}

  \draw[line width=2pt, black] (0,\yFiber) -- (1,\yFiber);

  \foreach \k in {1,...,\N} {

    \pgfmathsetmacro{\x}{rnd}

    \pgfmathparse{
      (
      (\x>=\a && \x<=\b)
      ||
      (\x>=\c && \x<=\d)
      ||
      (\x>=\e && \x<=\f)
      ) ? 1 : 0
    }

    \ifnum\pgfmathresult=1
      \draw[blueStrong, line width=1pt]
        (\x,\yFiber-\tickH) -- (\x,\yFiber+\tickH);
    \else
      \draw[blueFaint, line width=1pt]
        (\x,\yFiber-\tickH) -- (\x,\yFiber+\tickH);
    \fi
  }

  \draw[blue!35, line width=1.5pt]
    (\xm,\yFiber-\tickH)
    --
    (\xm,\yFiber+\tickH);

  \node[
    draw=blue,
    rounded corners=2pt,
    inner sep=3pt,
    below=20pt
  ]
  (scatterlabel)
  at (\xm,\yFiber-0.02)
  {$\{a_m,z_m\}$};

  \draw[
    blue,
    ->,
    line width=0.9pt
  ]
  (scatterlabel.north east)
  --
  (\xm+0.04,\yFiber-0.02);

  \node[below=8pt, xshift=-4pt]
    at (\a,\yFiber)
    {$z_{i-1}^{l}$};

  \node[below=8pt, xshift=6pt]
    at (\b,\yFiber)
    {$z_{i-1}^{r}$};

  \node[below=8pt, xshift=-4pt]
    at (\c,\yFiber)
    {$z_{i}^{l}$};

  \node[below=8pt, xshift=6pt]
    at (\d,\yFiber)
    {$z_{i}^{r}$};

  \node[below=8pt, xshift=-4pt]
    at (\e,\yFiber)
    {$z_{i+1}^{l}$};

  \node[below=8pt, xshift=6pt]
    at (\f,\yFiber)
    {$z_{i+1}^{r}$};

  \fill[black] (\mOne,\yFiber) circle (2.6pt);
  \fill[black] (\mTwo,\yFiber) circle (2.6pt);
  \fill[black] (\mThree,\yFiber) circle (2.6pt);

  \node[above=10pt]
    at (\mOne,\yFiber)
    {$z_{i-1}$};

  \node[above=10pt]
    at (\mTwo,\yFiber)
    {$z_i$};

  \node[above=10pt]
    at (\mThree,\yFiber)
    {$z_{i+1}$};

  \draw[->, line width=0.9pt]
    (0,\yArrowTop+0.3)
    to[bend left=22]
    (\mOne,\yArrowTop);

  \draw[->, line width=0.9pt]
    (\mOne,\yArrowTop)
    to[bend left=22]
    node[midway, above]
    {$\frac{T_S\,c}{2n_g}$}
    (\mTwo,\yArrowTop);

  \draw[->, line width=0.9pt]
    (\mTwo,\yArrowTop)
    to[bend left=22]
    node[midway, above]
    {$\frac{T_S\,c}{2n_g}$}
    (\mThree,\yArrowTop);

  \draw[->, line width=0.9pt]
    (\mThree,\yArrowTop)
    to[bend left=22]
    (1,\yArrowTop+0.3);

  \draw[
    green!60!black,
    dashed,
    line width=1pt
  ]
    (\c,-0.28)
    rectangle
    (\d,0.75);

  \node[
    green!60!black
  ]
    at (\mTwo, 1.02)
    {Scattering zone};


  \draw[
    decorate,
    decoration={
      brace,
      mirror,
      amplitude=5pt
    }
  ]
    (\a,\yBrace)
    --
    (\b,\yBrace)
    node[midway,below=6pt]
    {$\mathcal{E}_{i-1}$};

  \draw[
    decorate,
    decoration={
      brace,
      mirror,
      amplitude=5pt
    }
  ]
    (\c,\yBrace)
    --
    (\d,\yBrace)
    node[midway,below=6pt]
    {$\mathcal{E}_{i}$};

  \draw[
    decorate,
    decoration={
      brace,
      mirror,
      amplitude=5pt
    }
  ]
    (\e,\yBrace)
    --
    (\f,\yBrace)
    node[midway,below=6pt]
    {$\mathcal{E}_{i+1}$};

  \draw[line width=2pt, black]
    (0,\yTime)
    --
    (1,\yTime);

  \fill[black] (\mOne,\yTime) circle (2.6pt);
  \fill[black] (\mTwo,\yTime) circle (2.6pt);
  \fill[black] (\mThree,\yTime) circle (2.6pt);

  \node[below=6pt]
    at (\mOne,\yTime)
    {$t_{i-1}$};

  \node[below=6pt]
    at (\mTwo,\yTime)
    {$t_i$};

  \node[below=6pt]
    at (\mThree,\yTime)
    {$t_{i+1}$};

\end{tikzpicture}

\caption{
Random scatterers distributed along the fiber (upper black line) and the corresponding sampled phasors in the delay domain (lower black line). Blue ticks denote scatterers inside the scattering zones associated with the sampled phasors, whereas light-blue ticks indicate scatterers outside these zones.
}
\label{fig:scatters_distribution}
\end{figure}
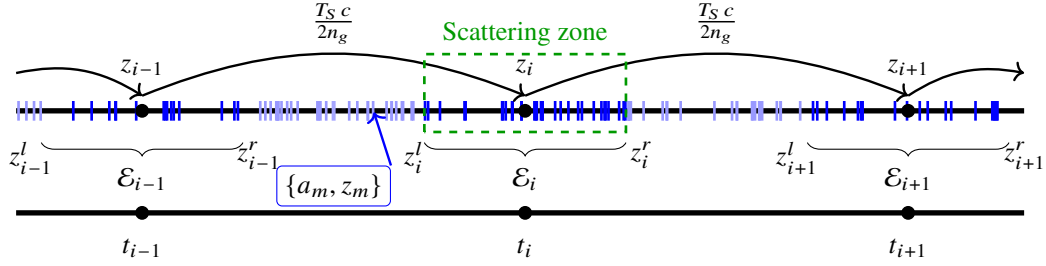

Due to the finite spatial extent of the probe pulse, not all scattering centers contribute to a given measurement. Instead, only the scatterers located within the corresponding \emph{scattering zone} participate in forming the measured backscattered signal, as illustrated in Fig.~\ref{fig:scatters_distribution}. The spatial length of this scattering zone is determined by the probe pulse duration and equals \(c T_{\mathrm{P}}/(2 n_g)\). The scattering zones themselves are centered in the vicinity of \(z_i\). In Fig.~\ref{fig:scatters_distribution}, blue ticks indicate scatterers inside the scattering zone, while light-blue ticks correspond to scatterers outside this region.

The optical phasor \(\mathcal{E}_i\) provides a convenient quantity for describing the coherently detected Rayleigh backscattered signal. On the one hand, it can be expressed in terms of the measurable amplitude and phase of the backscattered signal as \(\mathcal{E}_i = \sqrt{P_{\mathrm{RB}}}\,\bar{A}_i \exp(\mathrm{j}\,\bar{\phi}_i)\). On the other hand, the same quantity can be written as the coherent sum of contributions from the individual scatterers located within the \(i\)-th scattering zone bounded by \([z_i^{l}, z_i^{r}]\) (see Eq.~(7) in~\cite{main_model}):
\begin{equation}
\mathcal{E}_i 
=
\sum_{\; m:z_m \in [z_i^{l}, z_i^{r})}
a_m
\exp\!\left(
\mathrm{j}\,\frac{4\pi}{\lambda}
\int_{0}^{z_m} n(x)\,\mathrm{d}x
\right).
\label{eq:phasor_final}
\end{equation}

The average Rayleigh backscattered power associated with the \(i\)-th scattering zone can be related to the scattering amplitudes as
\[
\mathbb{E}\!\left[\sum_{m:z_m \in [z_i^{l}, z_i^{r})} |a_m|^2 \right] = P_{\mathrm{RB}},
\]
where the expectation \(\mathbb{E}\!\left[\right]\) is taken over different realizations of the random scatterer distribution.

The previously introduced left and right boundaries of the \(i\)-th scattering segment, \(z_i^{l}\) and \(z_i^{r}\), correspond to the round-trip delays \(t_i^{l}\) and \(t_i^{r}\) defined as
\begin{equation}
\begin{aligned}
t_i^{l,r} &= \frac{2}{c} \int_{0}^{z_i^{l,r}} n_g(x)\,\mathrm{d}x.
\end{aligned}
\end{equation}
Alternatively, the same delays can be written in terms of the sampling time \(t_i\) as
\begin{equation}
\label{eq:tau_i}
t_i^{l} = t_i - \frac{T_{\mathrm{P}}}{2}, \quad
t_i^{r} = t_i + \frac{T_{\mathrm{P}}}{2}.
\end{equation}

It is important to note that \(z_i\) does not necessarily coincide with the geometric midpoint of the segment \([z_i^{l}, z_i^{r}]\). Instead, it corresponds to the spatial location whose round-trip delay equals the temporal midpoint of the scattering zone.

In~\cite{main_model}, a dispersion-free model is adopted due to the relatively large pulse duration (typically $<10\,\mathrm{ns}$) used in fiber-optic sensing. Within this approximation, both the refractive index and the group refractive index are assumed to be frequency-independent and proportional to each other, such that they can be related by a constant factor $K_g = 1.01$~\cite{main_model}, i.e., $n_g(z) = K_g n(z)$, where $\lvert K_g - 1.01 \rvert < 10^{-3}$.

Following the same procedure as in~\cite{main_model}, we extract from Eq.~\eqref{eq:phasor_final} the common propagation path shared by all scatterers within the $i$-th scattering zone. This yields
\begin{eqnarray}
\label{eq:E_i_initial}
\mathcal{E}_i
&=&
\sum_{m:z_m \in [z_i^{l}, z_i^{r})}
a_m
\exp\!\left(
j\frac{4\pi}{\lambda}
\left(
\frac{1}{K_g}\int_{0}^{z_i}
\underbrace{K_g n(x)}_{n_g(x)}\,dx
+
\int_{z_i}^{z_m} n(x)\,dx
\right)
\right)
\nonumber \\
&=&
\exp\!\left(
j\,\frac{\omega}{K_g}
\underbrace{\frac{2}{c}\int_{0}^{z_i} n_g(x)\,dx}_{t_i}
\right)
\sum_{m:z_m \in [z_i^{l}, z_i^{r})}
a_m
\exp\!\left(
j\,\frac{4\pi}{\lambda}
\int_{z_i}^{z_i+\delta_{i,m}} n(x)\,dx
\right),
\end{eqnarray}
where $\delta_{i,m} = z_m - z_i$ denotes the distance between the reference position $z_i$ of the $i$-th scattering zone and the location of the individual scatterer $z_m$ within the same zone. Each scattering zone $i$ is characterized by its own set $\delta_{i,m}$, which determine the resulting amplitude and phase of the backscattered
signal. For simplicity, the index $i$ in $\delta_{i,m}$ is omitted and we write
$\delta_m$.

According to Eq.~\eqref{eq:E_i_initial}, the coordinate $z_i$ has a dual interpretation. On the one hand, as the upper limit of the first integral in the exponential term associated with the propagation delay, it represents the temporal midpoint of the scattering zone, $\tilde{z}_i$. On the other hand, as the part of upper limit of the integral in the coherent sum, $z_i+\delta_{i,m}$, it serves as a reference position within the scattering zone, $\check{z}_i$, relative to which the scatterer locations $z_m$ are defined.

\subsection{Model of Backscattered Signal Under Temperature Perturbation}\label{sec:dynamic_model}
\subsubsection{Model Assumptions}

To describe the slow-time evolution of the backscattered signal caused by thermal perturbations, we use the frame index $k$. Each probe pulse experiences a potentially different temperature distribution along the fiber, denoted by $T^k(z)$. Consequently, distributed quantities affected by temperature, such as the refractive index $n^k(z)$ and the group refractive index $n_g^k(z)$, may vary with the frame index $k$.

Due to thermo-optic and thermo-elastic effects, length-related quantities may also depend on \(k\). The temporal center of the scattering zone, \(\tilde{z}_i^{k}\), is determined by the group refractive index \(n_g^{k}(z)\), which in turn depends on the local temperature \(T^{k}(z)\), and therefore varies with \(k\). The reference position \(\check{z}_i^{k}\), being tied to a specific material point on the fiber, follows the corresponding change in fiber length, as do the relative scatterer positions \(\delta_{i,m}^{k}\).

In addition to the previously introduced assumptions of a dispersion-free model, noise-free operation, and neglected polarization effects, the following assumptions are adopted.

\begin{enumerate}

\item The spatial extent of the temperature-affected region is assumed to
span multiple consecutive scattering zones.

\item The boundaries of the scattering zones $z_i^l$ and $z_i^r$ are
assumed to be independent of the frame index $k$.

\end{enumerate}

Although temperature variations may cause small shifts of the scattering-zone boundaries $z_i^l$ and $z_i^r$, these shifts are on the order of micrometers, whereas the scattering-zone length $z_i^r - z_i^l = cT_P/(2n_g)$ is typically on the order of meters. Therefore, the backscattered signal originates from almost the same set of scatterers $\delta_{i,m}^k$, but observed from slightly different reference positions $\check{z}_i^k$ and $\tilde{z}_i^k$, which may vary with the frame index $k$.

Under these assumptions, the optical phasor of the backscattered signal can be written as
\begin{eqnarray}
\label{eq:E^k_i_2D}
\mathcal{E}_i^k
=
\exp\!\left(
j\,\frac{\omega}{K_g}
\underbrace{\frac{2}{c}\int_{0}^{\tilde{z}_i^k} n_g^k(x)\,dx}_{t_i}
\right)
\sum_{m:z^{k=0}_m \in [z_i^{l}, z_i^{r})}
a_m
\exp\!\left(
j\,\frac{4\pi}{\lambda}
\int_{\tilde{z}_i^k}^{\check{z}_i^k+\delta_m^k}
n^k(x)\,dx
\right).
\end{eqnarray}

It is important to note that the sampling time $t_i$ is determined by the ADC and therefore does not depend on the frame index $k$, even though its expression contains the $k$-dependent quantities $\tilde{z}_i^k$ and $n_g^k(z)$.

The coherent summation notation $z_m^{k=0}$ indicates that the set of scatterers included in the summation is defined by the scattering zone in the reference frame ($k=0$).

The total propagation path represented by the two integrals in Eq.~\eqref{eq:E^k_i_2D} extends to the temperature-dependent scatterer position $z_m^k = \check{z}_i^k + \delta_m^k$, which corresponds to a specific material point of the fiber and is therefore affected by the thermo-elongation effect.

\subsubsection{Definition of a Temperature Event}

The local temperature evolution $T(t,z)$, shown schematically in Fig.~\ref{fig:measurements}, is in principle a continuous function of time and spatial coordinate. Ideally, this evolution would be monitored continuously. In practice, however, continuous monitoring is often not possible due to limitations in data storage and processing.

Instead, the signal is acquired in short recording intervals referred to as \emph{measurement cycles}. Each measurement cycle consists of $K_f + 1$ consecutive frames and spans a total duration of $(K_f + 1)T_R$. Successive measurement cycles are separated by longer \emph{inter-cycle intervals}. These inter-cycle intervals effectively determine the sampling rate of the macroscopic temperature evolution $T(t)$ and must be chosen sufficiently small to ensure that its essential temporal behaviour can be reconstructed.

In $\phi$-OTDR systems, the measurement principle is inherently differential, since the detected phase is evaluated relative to a reference frame. Consequently, the system is sensitive to temperature variations rather than to the absolute temperature. The temperature change relative to the reference frame is defined as
\begin{equation}
\Delta T^k(z) = T^k(z) - T^0(z),
\end{equation}
where $T^0(z)$ denotes the temperature distribution in the reference
frame.

A temperature event is considered to occur if there exists a position $z$ for which $\Delta T^k(z) \neq 0$. The event type is then classified according to the sign of the temperature change:
\begin{equation}\label{eq:def_T_event}
\operatorname{sign}\!\big[\Delta T^k(z)\big] =
\begin{cases}
+1, & \text{heating at } z, \\
0, & \text{no event}, \\
-1, & \text{cooling at } z.
\end{cases}
\end{equation}

\subsubsection{Shift of the Temporal Midpoint Due to the Thermo-Optic Effect}

We consider a temperature-affected fiber segment at position $z$, where the temperature evolves from an initial distribution $T^0(z)$ (reference frame) to $T^k(z)$ in the $k$-th measurement frame. A change in fiber temperature induces proportional changes in both the phase refractive index $n(T)$ and the group refractive index $n_g(T)$. Consequently, the spatial and temporal dependence of $T^k(z)$ is directly reflected in the refractive-index distributions via the thermo-optic effect:
\begin{eqnarray}
   n^k(z) &=& n^{0}(z) + \alpha_n \,\Delta T^{k}(z), \\ 
   n_g^k(z) &=& n_g^0(z) + \alpha_g \,\Delta T^{k}(z)
   \;=\; K_g n^0(z) + K_g \alpha_n \,\Delta T^{k}(z),
\label{eq:thermo-optic effect}
\end{eqnarray}
where $\alpha_n$ is the thermo-optic coefficient, taken as $1.178 \times 10^{-5}~\mathrm{K}^{-1}$~\cite{ref12}, and $\Delta T^k(z) = T^k(z) - T^0(z)$, with $\Delta T^{k=0} \triangleq 0$.

The temperature perturbation $\Delta T^k(z)$ modifies the group refractive index distribution and thus the local group velocity of the optical pulse. In the case of heating, $\Delta T^k(z) > 0$, the increased group refractive index reduces the pulse velocity. As a result, for a fixed detection delay, the effective temporal midpoint of the corresponding scattering zone shifts towards smaller spatial coordinates. 

The round-trip delay associated with the $i$-th sample must satisfy
\begin{equation}
\label{eq:delay_0-k_frame}
t_i
= \frac{2}{c} \int_{0}^{z_i^{0}} n_g^{0}(x)\,dx
= \frac{2}{c} \int_{0}^{\tilde{z}_i^{k}} n_g^{k}(x)\,dx = i\,T_S.
\end{equation}

Substituting the thermo-optic relation from Eq.~\eqref{eq:thermo-optic effect} into the second integral of Eq.~\eqref{eq:delay_0-k_frame}, the delay in frame $k$ can be written as
\begin{equation}
\label{eq:first_part_eq}
t_{i}
= \frac{2}{c}
\left[
\int_{0}^{\tilde{z}^{k}_{i}} n_{g}^{0}(x)\,dx
+ \alpha_g \int_{0}^{\tilde{z}^{k}_{i}} \Delta T^k(x)\,dx
\right].
\end{equation}

On the other hand, by definition \(t_i = iT_{\mathrm{S}}\) is independent of the frame index \(k\). For the initial frame, \(k = 0\), it can be decomposed via \(\tilde{z}_i^{k}\) as
\begin{equation}
\label{eq:second_part_eq}
t_{i}
= \frac{2}{c}
\left[
\int_{0}^{\tilde{z}^{k}_{i}} n_{g}^{0}(x)\,dx
+ \int_{\tilde{z}^{k}_{i}}^{z^{0}_{i}} n_{g}^{0}(x)\,dx
\right].
\end{equation}

Equating the right-hand sides of Eqs.~\eqref{eq:first_part_eq} and~\eqref{eq:second_part_eq} and cancelling identical terms yields
\begin{equation}
\label{eq:temp_refrac0}
\int_{\tilde{z}^{\,k}_{i}}^{z^0_{i}}
n_{g}^{0}(x)\,dx = 
\alpha_g
\int_{0}^{\tilde{z}^{\,k}_{i}}
\Delta T^{k}(x)\,dx =\alpha_g \left( \int_{0}^{z^{0}_{i}} \Delta T^{k}(x)\,dx + \int_{z^0_i}^{\tilde{z}^{\,k}_{i}}\Delta T^{k}(x)\,dx\right),
\end{equation}
where the expression after the second equality is gotten by splitting up the distance from interrogator to $\tilde{z}^k_i$ through the initial position $z^0_i$.

The both side of Eq.~\eqref{eq:temp_refrac0} can be further simplified. Since the displacement between $\tilde{z}^{\,k}_{i}$ and $z^0_{i}$ is typically on the order of a few micrometres, it is reasonable to approximate $n_g^0(z)$ as well as $\Delta T^{k}(z)$ as constants over this interval. Taking $n_g^0(z_i^0) = K_g n_i^0$ ($n^0_i = n^0(z^0_i)$) from left part and $\Delta T^{k}(z^0_i)=\Delta T^{k}_i$ from right side outside the integral we end up with
\begin{equation}
\label{eq:some_in_middle}
K_{g} n^0_i \left(z^0_i - \tilde{z}^k_i\right) = K_g \,\alpha_n \left( \int_{0}^{z^{0}_{i}} \Delta T^{k}(x)\,dx - \Delta T^{k}_i\left(z^0_i - \tilde{z}^k_i\right) \right),
\end{equation}
canceling out $K_g$ and neglecting the small contribution of $\alpha_n \Delta T^k_i$ relative to $n^0_i$ leads to an explicit expression for the shifted scattering-zone centre:
\begin{equation}
\label{eq:shift_thermo-optic_final}
    \tilde{z}^{k}_{i}
    \;\approx\; z^0_i - \frac{\alpha_n}{n^0_i} \int_{0}^{z^0_i} \Delta T^k(x)\,dx.
\end{equation}

Eq.~\eqref{eq:shift_thermo-optic_final} shows that the scattering-zone centre shifts linearly with the cumulative temperature change, with a sign opposite to that of $\Delta T^k$.

\subsubsection{Shift of the Reference Position Due to Thermo-Expansion}

Recall that $z_i$ denotes the reference position along the fiber relative to which the locations of scatterers within the $i$-th resolution cell are defined. Due to thermal expansion of the fiber, this reference position is displaced to a new position $\check{z}_i^{\,k}$ in the $k$-th measurement frame. The thermo-expansion-induced shift can be expressed as
\begin{equation}
\check{z}_i^{\,k}
= z_i^{0} + \alpha_d \int_{0}^{z_i^{0}} \Delta T^{k}(x)\,dx,
\label{eq:thermo-expansion}
\end{equation}
where $\alpha_d$ is the linear thermal expansion coefficient of the fiber, taken as $0.55 \times 10^{-6}~\mathrm{K}^{-1}$~\cite{ref12}.

By comparing Eq.~\eqref{eq:thermo-expansion} with the thermo-optic shift of the scattering-zone centre given in Eq.~\eqref{eq:shift_thermo-optic_final}, it is evident that the reference position $\check{z}_i^{\,k}$ and the temporal midpoint $\tilde{z}_i^{\,k}$ shift in opposite directions in response to temperature variations. At the reference frame ($k=0$), where $\Delta T^0(z)=0$, both positions coincide, i.e., $\check{z}_i^{0} = \tilde{z}_i^{0} = z_i^{0}$.

\begin{figure}[t]
    \centering\includegraphics[width=13cm]{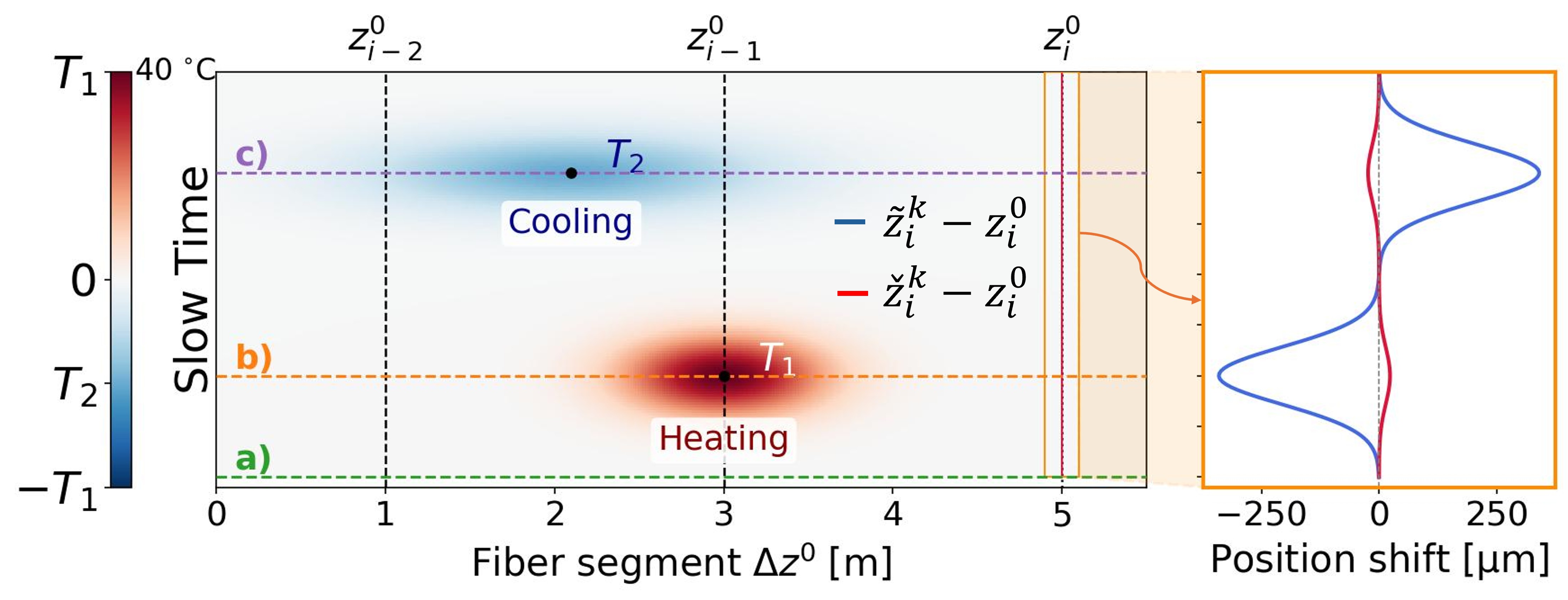}
    \centering\includegraphics[width=13cm]{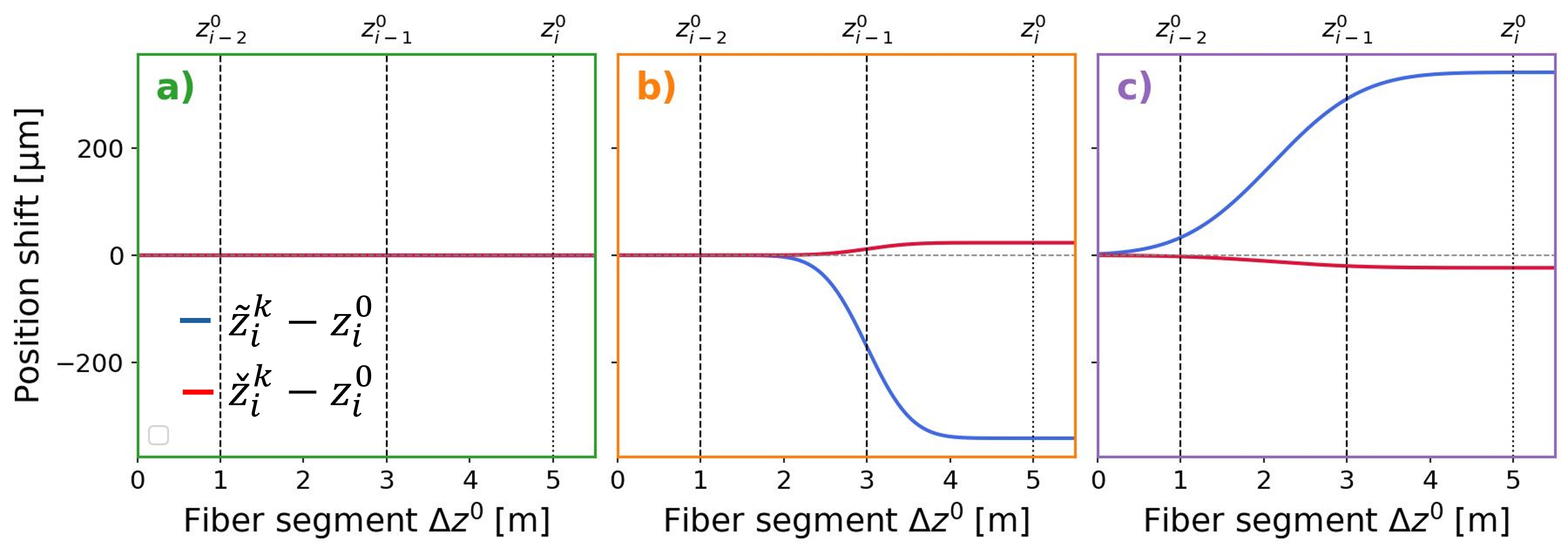}
    \caption{(Top) Temperature perturbation along the fiber in slow time and the resulting shifts of the logical \((\tilde{z}_i^{k})\) and material \((\check{z}_i^{k})\) positions after the temperature event. (Bottom) Spatial distribution of the logical \((\tilde{z}_i^{k})\) and material \((\check{z}_i^{k})\) position shifts along the fiber at different slow-time instants.}
    \label{fig:theoretical}
\end{figure}

It is important to clearly distinguish between these two positions. The center of the scattering zone, \(\tilde{z}_i^{\,k}\), is a \emph{logical} position defined by the optical round-trip delay and may therefore shift along the fiber when the group refractive index changes. In contrast, the reference position, \(\check{z}_i^{\,k}\), corresponds to a \emph{material} point on the fiber and moves together with the glass due to thermal expansion. This difference in behavior is illustrated in Fig.~\ref{fig:theoretical}. In the top panel, a heating event centered near \(z_{i-1}^0\), followed in slow time by a cooling event between \(z_{i-2}^0\) and \(z_{i-1}^0\), produces distinct shifts of \(\check{z}_i^{\,k}\) and \(\tilde{z}_i^{\,k}\) in the vicinity of \(z_i\) and at subsequent positions \(z_{\hat{i}} > z_i\). Both events are modeled as 2D Gaussians in the slow-time--distance plane, with their principal axes aligned with slow time and fiber position; the heating event has twice the amplitude but half the spatial width along \(z\). Owing to the integral nature of both \(\check{z}_i^{\,k}\) and \(\tilde{z}_i^{\,k}\), the net shift at position \(z_i\) is nearly identical in magnitude but opposite in sign. The distributions of the shifts \(\tilde{z}_i^{\,k} - z_i^0\) and \(\check{z}_i^{\,k} - z_i^0\) along the reference coordinate \(z^0\) for the first frame are shown in the bottom panel of Fig.~\ref{fig:theoretical} for several slow-time instants. As can be seen, the temperature variation manifests primarily through changes in the logical position \(\tilde{z}_i^{\,k}\), whereas the thermo-elongation captured by \(\check{z}_i^{\,k}\) plays a secondary role.

The temperature-dependent offset between the reference position and the scattering-zone centre follows directly from Eqs.~\eqref{eq:thermo-expansion} and~\eqref{eq:shift_thermo-optic_final}:
\begin{equation}
    \check{z}_i^{\,k} - \tilde{z}_i^{\,k} =
    \left(
        \alpha_d + \frac{\alpha_n}{n_i^{0}}
    \right)
    \int_{0}^{z_i^{0}} \Delta T^{k}(x)\,dx.
    \label{eq:distance}
\end{equation}

Eq.~\eqref{eq:distance} highlights that the separation between $\check{z}_i^{\,k}$ and $\tilde{z}_i^{\,k}$ is not determined solely by the local temperature change in the vicinity of $z_i$, but rather by the cumulative temperature variation along the fiber segment extending from the interrogator to position $z_i$. Consequently, any temperature perturbation occurring at positions $z_{i^{*}} < z_i$ contributes to the observed offset.

The scatterer positions $z_m^{\,k}$ are material points embedded in the fiber and therefore undergo thermo-expansion in the same manner as the reference position in Eq.~\eqref{eq:thermo-expansion}. Accordingly, the thermo-expansion-induced displacement of a scatterer can be described by an expression analogous to Eq.~\eqref{eq:thermo-expansion}, with the index $i$ replaced by $m$.
The position of each scatterer relative to reference position $\check{z}_i^{\,k}$ in the $k$-th measurement frame is defined as
\begin{eqnarray}
    \label{eq:sc}
    \delta_m^{\,k}
    &=&
    z_m^{\,k} - \check{z}_i^{\,k} 
    \\ \nonumber
    &=& 
    z_m^{\,0} - \check{z}_i^{\,0} + \alpha_d \left(\int_0^{z_m^{\,0}} \Delta T^k(x) \,dx - \int_0^{\check{z}_i^{\,0}} \Delta T^k(x) \,dx \right)
    =
    \delta_m^{\,0}
    +
    \alpha_d
    \int_{\check{z}_i^{0}}^{z_m^{0}}
    \Delta T^{k}(x)\,dx,
\end{eqnarray}
where $\delta_m^{\,0} = z_m^{\,0} - \check{z}_i^{\,0}$ denotes the relative scatterer position in the reference frame. 

The temperature perturbation $\Delta T^k(z)$ varies smoothly along the fiber.
If its spatial correlation length is much larger than the scattering-zone width $cT_P/(2n_g)$, it can be locally approximated by a first-order Taylor expansion:
\begin{equation}
    \Delta T^k(z_i^0 + \delta_m^0)
    \approx
    \Delta T^k(z_i^0)
    +
    \left.
    \frac{d \Delta T^k}{dz}
    \right|_{z_i^0}
    \delta_m^0,
    \qquad
    z \in [z_i^l, z_i^r],
    \label{eq:taylor_expansion}
\end{equation}
where $\left.\frac{d \Delta T^k}{dz}\right|_{z_i^0}$ denotes the spatial temperature gradient evaluated at position $z_i^0$. Substituting Eq.~\eqref{eq:taylor_expansion} into Eq.~(\ref{eq:sc}) and carrying out the integration yields the following explicit expression for the thermo-expansion-induced shift of the scatterer position:
\begin{equation}
    \delta_m^{\,k}
    =
    \delta_m^{\,0}
    \left(
    1
    +
    \alpha_d \Delta T_i^{k}
    +
    \frac{1}{2}\alpha_d
    \left.
    \frac{d\Delta T^{k}}{dz}
    \right|_{z_i^{0}}
    \delta_m^{\,0}
    \right).
\label{eq:scatter_shift_final}
\end{equation}

\subsubsection{Local Temperature Sensitivity of the Rayleigh Backscattered Phasor}

Based on the introduced variables, Eq.~\eqref{eq:E^k_i_2D} can be rewritten as
\begin{eqnarray}
\label{eq:E^k_i}
\mathcal{E}_i^k
&=&
\exp\!\left( j\,\frac{\omega}{K_g}\,t_i \right)
\sum_{m:z^{k=0}_m \in [z_i^{l}, z_i^{r})}
a_m\,
\exp\!\left(
j\,\frac{4\pi}{\lambda}
\left[
\int_{\tilde{z}^k_i}^{\check{z}^k_i} n^{k}(x)\,dx
+
\int_{\check{z}^k_i}^{\check{z}^k_i+\delta^k_m} n^{k}(x)\,dx
\right]
\right)\\
&=&
\exp\!\left( j\,\frac{\omega}{K_g}\,t_i \right)
\exp\!\left( j\,\frac{4\pi}{\lambda} \int_{\tilde{z}^k_i}^{\check{z}^k_i} n^{k}(x)\,dx \right)
\sum_{m:z^{k=0}_m \in [z_i^{l}, z_i^{r})}
a_m\,
\exp\!\left( j\,\frac{4\pi}{\lambda} \int_{\check{z}^k_i}^{\check{z}^k_i+\delta^k_m} n^{k}(x)\,dx \right)\nonumber.
\end{eqnarray}

Here, the optical phase integral from the scattering-zone centre $\tilde{z}_i^k$ to an individual scatterer position is split at the (thermally shifted) reference position $\check{z}_i^k$, yielding a common term $\int_{\tilde{z}^k_i}^{\check{z}^k_i} n^{k}(x)\,dx$ and a scatterer-dependent term $\int_{\check{z}^k_i}^{\check{z}^k_m+\delta^k_m} n^{k}(x)\,dx$. Since the former does not depend on the scatterer index $m$, it can be factored out of the summation, leading to the second line of Eq.~\eqref{eq:E^k_i}.

The common exponential term in Eq.~\eqref{eq:E^k_i} arises exclusively from temperature-induced effects and equals unity in the reference frame ($k=0$). This follows directly from the fact that, by definition, the scattering-zone centre and the reference position coincide in the reference frame, i.e.,$\int_{\tilde{z}_i^{k=0}}^{\check{z}_i^{k=0}} n^{0}(x)\,dx = 0$.
For subsequent frames, this term introduces an additional phase shift that is directly observable in the coherently detected signal.

To analyse this contribution in more detail, we account for the thermo-optic effect described in Eq.~\eqref{eq:thermo-optic effect}. Since the reference position $\check{z}_i^{\,k}$ and the scattering-zone centre $\tilde{z}_i^{\,k}$ are separated by a small distance, both the refractive index and the temperature variation can be approximated as constant over this interval. Using the previously derived separation $\check{z}_i^{\,k} - \tilde{z}_i^{\,k}$ from Eq.~\eqref{eq:distance}, the corresponding phase contribution can be written as

\begin{eqnarray}
\label{eq:main_part}
\frac{4\pi}{\lambda}
\int_{\tilde{z}_i^{\,k}}^{\check{z}_i^{\,k}} n^{k}(x)\,dx
&\approx&
\underbrace{K_{\alpha} \int_{0}^{z_i^0} \Delta T^{k}(x)\,dx}_{\text{A}}
+\,\,
\underbrace{
\frac{\alpha_n}{n^0_i}\,  \Delta T_i^{k} \,
K_{\alpha}
\int_{0}^{z_i^0} \Delta T^{k}(x)\,dx}_{\text{B}},
\end{eqnarray}
where $K_{\alpha} = \frac{4\pi}{\lambda} \left( \alpha_d n_i^{0} + \alpha_n \right)$ is a coefficient that remains constant within a single measurement cycle, but may vary slowly between cycles due to the temperature dependence of the refractive index $n_i^{0}$.

Term (A) represents the dominant linear contribution arising from the combined thermo-optic and thermo-expansion effects, while term (B) corresponds to a higher-order correction proportional to the local temperature change. Although term (B) is multiplied by the small factor $\alpha_n \Delta T_i^{k}$, its contribution appears in the argument of a complex exponential and therefore must be evaluated quantitatively.

For a representative temperature rate of $1~\mathrm{K/s}$ and a temperature-affected fiber segment of $25~\mathrm{m}$, the accumulated phase associated with term (B) over a single measurement cycle is on the order of $10^{-7}$–$10^{-5}$~rad. In contrast, the phase contribution of term (A) is on the order of $10$~rad. Consequently, term (B) can be safely neglected under slow temperature-rate conditions, while term (A) dominates the temperature-induced phase evolution.

In a similar manner, and under the additional reasonable assumption that the spatial derivative of the refractive index vanishes locally,
$\left.\frac{dn^k}{dz}\right|_{z_i^{0}} = 0$,
the scatterer-dependent term in Eq.~\eqref{eq:E^k_i} can be analysed. Neglecting all terms that are quadratic in the thermo-expansion and thermo-optic coefficients $\alpha_d$ and $\alpha_n$, the corresponding phase contribution can be approximated as
\begin{equation}
\label{eq:external_part}
\frac{4\pi}{\lambda}
\int_{\check{z}_i^{\,k}}^{\check{z}_i^{\,k}+\delta_m^{\,k}} n^{k}(x)\,dx
=
\delta_m^{\,0}
\left(
\frac{4\pi}{\lambda} n_i^{0}
+
K_{\alpha}\,\Delta T_i^{k}
+
\frac{1}{2}
K_{\alpha}
\left.
\frac{d\Delta T_i^{k}}{dz}
\right|_{z_i^{0}}
\delta_m^{\,0}
\right),
\end{equation}

By combining Eqs.~\eqref{eq:main_part} and~\eqref{eq:external_part} with the phasor representation in Eq.~\eqref{eq:E^k_i}, the resulting expression for the Rayleigh backscattered phasor becomes
\begin{eqnarray}
\label{eq:phasor_end}
\mathcal{E}_i^k
&=&
\exp\!\left( j\,\frac{\omega}{K_g}\,t_i \right)
\exp\!\left( j\,K_{\alpha} \int_{0}^{z_i^0} \Delta T^{k}(x)\,dx \right)
\nonumber\\
&&\times
\sum_m a_m
\exp\!\left( j\,\frac{4\pi}{\lambda} n_i^0 \delta_m^{\,0} \right)
\exp\!\left( j\,K_{\alpha}\,\Delta T_i^{k}\,\delta_m^{\,0} \right)
\exp\!\left(
j\,K_{\alpha}
\left.
\frac{d\Delta T_i^{k}}{dz}
\right|_{z_i^{0}}
(\delta_m^{\,0})^2
\right).
\end{eqnarray}

Eq.~\eqref{eq:phasor_end} explicitly demonstrates how temperature variations modify the Rayleigh backscattered phasor. In comparison to the initial phasor
\[
\mathcal{E}_i
=
\exp\!\left( j\,\frac{\omega}{K_g}\,t_i \right)
\sum_m a_m
\exp\!\left( j\,\frac{4\pi}{\lambda} n_i^0 \delta_m^{\,0} \right),
\]
defined in Eq.~\eqref{eq:E_i_initial}, temperature-induced effects manifest primarily as additional phase factors, most notably
\(
\exp\!\left( j\,K_{\alpha} \int_{0}^{z_i^0} \Delta T^{k}(x)\,dx \right),
\)
as well as weaker amplitude and phase modifications arising from the local temperature change $\Delta T_i^k$ and its spatial gradient, as indicated in Eq.~\eqref{eq:phasor_end}.

Taking the argument of Eq.~\eqref{eq:phasor_end}, the resulting phase evolution can be expressed as
\begin{equation}
\label{eq:phase_end}
\phi_i^{k}
=
\phi_i^{0}
+
\delta\phi_i^{k}
+
K_{\alpha}
\int_{0}^{z_i^0} \Delta T^{k}(x)\,dx,
\end{equation}
where
\begin{equation}
\label{eq:iitial_phase}
\phi_i^{0}
=
\frac{\omega}{K_g}\,\tau_i
+
\tan^{-1}
\!\left(
\frac{\sum_m a_m \sin\!\left(\frac{4\pi}{\lambda_0} n_i^0 \delta_m^{\,0}\right)}
     {\sum_m a_m \cos\!\left(\frac{4\pi}{\lambda_0} n_i^0 \delta_m^{\,0}\right)}
\right)
\end{equation}
denotes the random initial phase at the reference temperature $T_i^{0}$. This phase remains constant over the slow-time domain, i.e., it does not depend on the frame index $k$. The term $\delta\phi_i^{k}$ represents a small phase perturbation induced by the local temperature variation $\Delta T_i^{k}$ and its spatial gradient
$\left.\frac{d\Delta T_i^{k}}{dz}\right|_{z_i^{0}}$, and varies weakly with slow time.

Since thermal processes evolve much more slowly than optical measurements, the temperature change can be approximated as
\begin{equation}
\label{eq:temperature_rate}
\Delta T^k(z)
\approx
\dot{T}(z)\,k\,T_R,
\end{equation}
where $\dot{T}(z)=\frac{dT}{dt}(z)$ denotes the local temperature rate over slow-time domain, assumed to be constant over the interval of interest. Substituting Eq.~\eqref{eq:temperature_rate} into Eq.~\eqref{eq:phase_end} yields
\begin{equation}
\label{eq:phase_measurable_quality}
\phi_i^{k}
=
\phi_i^{0}
+
\delta\phi_i^{k}
+
K_{\alpha}
\left(
\int_{0}^{z_i^0} \dot{T}(x)\,dx
\right)
k\,T_R.
\end{equation}

Eq.~\eqref{eq:phase_measurable_quality} establishes a direct link between the measured phase evolution and the underlying temperature dynamics. In the slow-time domain, the phase evolution can be approximated by a linear model,
\(
\phi_i^{k}
\approx
\phi_i^{0}
+
K_{\alpha} K_i\, k\,T_R,
\)
where $\delta\phi_i^{k}$ is treated as a noise term. The slope parameter
\begin{equation}
\label{eq:tangent_coef}
K_i
=
\int_{0}^{z_i^0} \dot{T}(x)\,dx
\end{equation}
represents the cumulative temperature rate along the fiber segment extending from the interrogator to position $z_i^0$.

It should be noted that the coefficient $K_{\alpha}$ depends weakly on the
spatial coordinate through the refractive index $n_i^0$, i.e.,
$K_{\alpha} = K_{\alpha}(z_i)$. In principle, this position dependence
could be taken into account in the temperature-reconstruction algorithm.
However, the corresponding variation of $n_i^0$ along a standard
single-mode fiber is typically very small, and therefore the spatial
variation of $K_{\alpha}$ is neglected in the present analysis for
simplicity.

It is worth noting that coefficients of the form $K_{\alpha}$, linking
phase variations to temperature perturbations, have been previously
introduced in the literature~\cite{morris}. However, to the best of
our knowledge, the complete relationship between the local temperature
distribution and the Rayleigh backscattered signal derived in
Eq.~\eqref{eq:phasor_end} has not been previously reported.

\section{Experiment}\label{sec:experiment}
\subsection{$\phi$-OTDR Setup for Qualitative Validation of Distributed Temperature Sensing}

\begin{figure}[b]
  \centering
  \includegraphics[width=13cm]{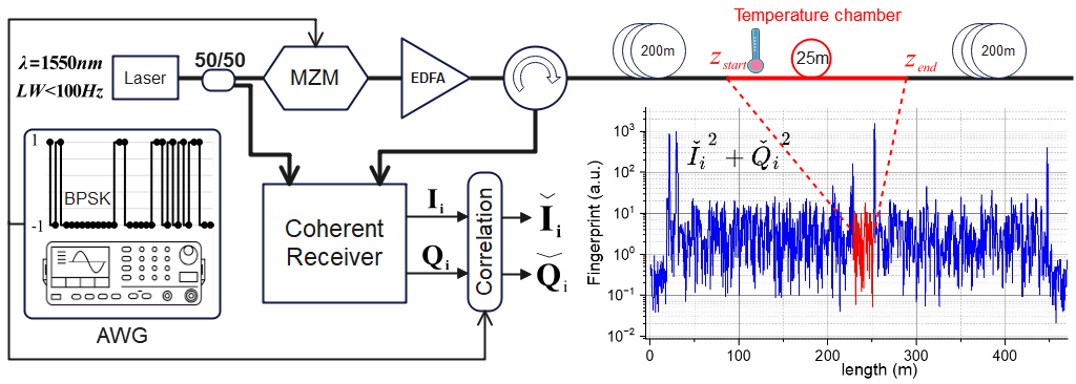}
  \caption{$\phi$-OTDR experimental setup for quantitative distributed temperature validation.}
  \label{fig:OTDR_setup}
\end{figure}

The experimental setup employed for $\phi$-OTDR measurements and investigation of temperature effects on Rayleigh backscattering is schematically illustrated in Fig.~\ref{fig:OTDR_setup}. The output of a highly coherent continuous-wave (CW) laser (linewidth $< 100$~Hz, 1550~nm) is split into two paths using a beam splitter. One portion of the optical signal serves as the local oscillator (LO), while the remaining part is utilized to probe the fiber under test (FUT).
The probing signal passes through a Mach-Zehnder modulator (MZM), driven at a bit rate of 125~Mbit/s by an arbitrary waveform generator (AWG), modulating the optical signal via binary phase-shift keying (BPSK). This modulation produces optical pulses of 160~cm in length, corresponding to a scattering zone of 80~cm. The AWG generates a probe sequence consisting of a 4096-bit pseudo-random binary sequence, followed by zero-padding of 5000 symbols, resulting in a total frame duration $T_{R}$ of 72.8~$\mu$s.

Following modulation, the optical signal is amplified by an Erbium-Doped Fiber Amplifier (EDFA) before being launched into the FUT through an optical circulator. The FUT comprises two 200~m-long lead-in and termination fibers interconnected by a 25~m patch cord, which acts as the sensing fiber segment. The backscattered and reflected optical signals from the FUT are coherently combined with the LO at the input of a coherent receiver, employing a self-homodyne detection scheme to extract amplitude and phase information. A real-time oscilloscope records the four optical field components at a sampling rate of 625~MS/s for subsequent signal processing and analysis. The first stage of digital signal processing applies a correlation operation, which converts the response to the transmitted pulse sequence into the equivalent response of a single optical pulse~\cite{ref18,ref19}. The output of the correlation block, denoted as $\check{I}_i^{k}+j\check{Q}_i^{k}$, represents the single-pulse response and corresponds to $I_i^{k}+jQ_i^{k}$ in Eq.~\eqref{eq:I+Q}. The temperature-affected $25\,\mathrm{m}$ fiber patch cord is placed inside a temperature-controlled chamber, as illustrated in Fig.~\ref{fig:OTDR_setup}. An example fingerprint of a single measurement frame, defined as
\begin{equation}
\label{eq:fingerprint}
    FP_i^{k} = (\check{I}_i^{k})^{2} + (\check{Q}_i^{k})^{2},
\end{equation}
is shown beneath the FUT and represents a characteristic spatial signature of the sensing fiber.

\section{Results}\label{sec:results}

\subsection{Algorithm of Event Detection}

\begin{figure}[b]
\centering
\includegraphics[width=13cm]{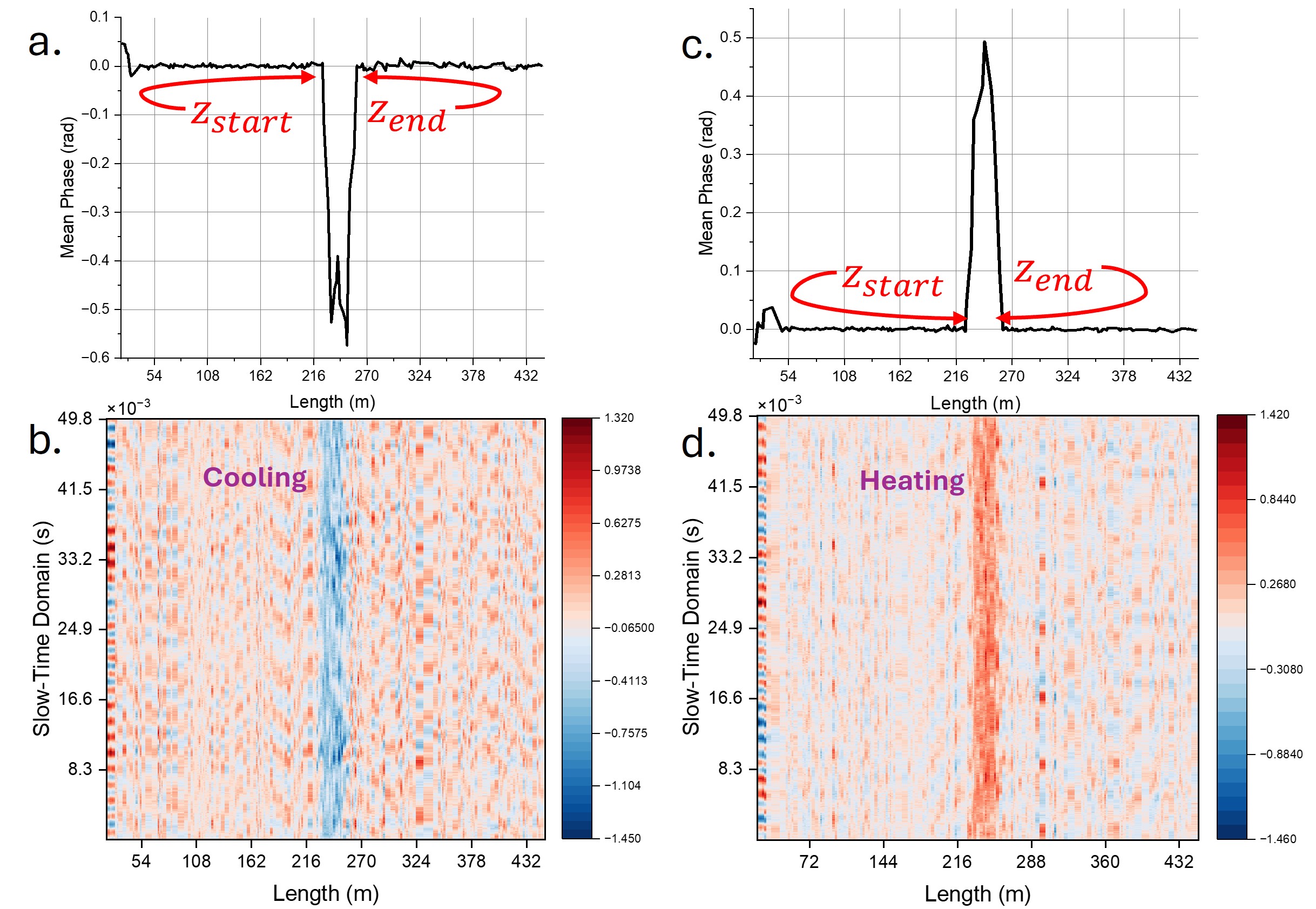}
\caption{Evolution of the second-order phase differential function $\Delta \psi_{i,m}^k$ over the slow-time domain ($k$) during heating (b) and cooling (d) processes. Subfigures (a) and (c) show the corresponding averages over the slow-time domain for the heating and cooling cases, respectively.}
\label{fig:phase}
\end{figure}

The first step in processing the measured data for the extraction of useful information is the detection of temperature events. This task consists of two main objectives: (i) identifying the presence of a temperature event, and (ii) determining its spatial location along the fiber.

The theoretical framework developed in the previous sections assumes a noiseless measurement scenario. In practice, however, experimental data are inevitably affected by various noise sources. A comprehensive analysis of noise mechanisms is beyond the scope of the present work; instead, we focus on general and robust concepts relevant for temperature event detection.

Additive Gaussian noise predominantly affects low-power regions of the Rayleigh backscattered signal, leading to unreliable phase estimates in the slow-time domain. To mitigate this effect, the Rayleigh backscattered signal power, represented by the fingerprint $FP_i^{k=0}$, is used as a quality metric. As a first step in the temperature event-detection algorithm, only fiber positions $i$ for which the fingerprint exceeds an experimentally determined threshold $Th$ ($FP_i^{k=0} > Th$). This step effectively excludes low-SNR regions from further processing.

The slow-time evolution of the absolute phase $\phi_i^k$ at a given fiber position
$i$ is additionally governed by laser phase noise, which was neglected in the model
derivation. This noise manifests through the term
$\mathrm{e}^{\mathrm{j}(\phi_{\mathrm{RB}} - \phi_{\mathrm{LO}})}$ in
Eq.~\eqref{eq:I+Q}, where the quantity $(\phi_{\mathrm{RB}} - \phi_{\mathrm{LO}})(t)$
acquires a slow time dependence. Although the rate of phase drift is small due to
the high coherence of the laser, the accumulated phase noise over the full fiber
length can become significant. It is therefore preferable to work with phase
differences rather than absolute phase values, as the common-mode phase drift
cancels upon subtraction. Accordingly, we define the differential phase between
two fiber positions $i$ and $m$ as
\begin{equation}
\label{eq:phase_difference}
    \psi_{i,m}^k = \operatorname{unwrap}\!\left(\phi_i^k - \phi_m^k\right),
\end{equation}
where the unwrapping operation is performed along the slow-time index $k$.

For practical signal processing, it is convenient to further consider the slow-time phase increment of the phase difference, defined as $\Delta \psi_{i,m}^k
= \psi_{i,m}^k - \psi_{i,m}^{k-1}$.
Using the phase–temperature relation derived above, this yields
\begin{equation}
\label{eq:unwrapped}
    \Delta \psi_{i,m}^k \approx K_{\alpha}\,T_R \int_{z_m^0}^{z_i^0} \dot{T}(x)\,dx.
\end{equation}

The slow-time phase increment $\Delta \psi_{i,m}^k$ is directly proportional to the cumulative temperature rate between positions $z_m^0$ and $z_i^0$. For a nonzero temperature rate, $\Delta \psi_{i,m}^k$ remains constant with respect to the frame index $k$, while its sign is determined by the sign of $\dot{T}(x)$ over the interval $[z_m^0, z_i^0]$.

Fig.~\ref{fig:phase} illustrates the slow-time evolution of $\Delta \psi_{i,m}^k$ for the case where $\check{i}=\mathrm{findpeak}(FP_i^0 > Th)$ and $m=\check{i}-1$. As can be observed in Fig.~\ref{fig:phase}(b) and (d), heating and cooling events are clearly distinguishable in the raw slow-time phase increments. Averaging over the slow-time index yields the results shown in Fig.~\ref{fig:phase}(a) and (c), respectively. Pronounced positive peaks are associated with heating events, whereas negative peaks correspond to cooling events, in accordance with the temperature event definition given in Eq.~\eqref{eq:def_T_event}. These features can therefore be used for reliable temperature-event detection, event-type classification, and estimation of the event boundaries $z_{\mathrm{start}}$ and $z_{\mathrm{end}}$.

\subsection{Algorithm of temperature change recovering}

Quantitative analysis is performed by applying a linear approximation in the slow-time domain $k$ to the phase difference defined in Eq.~\eqref{eq:phase_difference}. Substituting Eq.~\eqref{eq:phase_measurable_quality} into this expression yields
\begin{equation}
\label{eq:last_one}
\psi_{i,m}^k
=
\psi_{i,m}^0
+
\delta \psi_{i,m}^k
+
K_{\alpha}
\int_{z_m^0}^{z_i^0}
\dot{T}(x)\,dx\;
k\,T_R .
\end{equation}

To ensure reliable estimation of the temperature rate, the coefficient $K_i$ form Eq.~\eqref{eq:tangent_coef} must be sufficiently large. This can be achieved by selecting the reference position $z_m^0 < z_{\mathrm{start}}$, such that the integration interval covers the entire temperature-affected region $[z_{\mathrm{start}}, z_i^0]$.

\begin{figure}[t!]
\centering
\includegraphics[width=13cm]{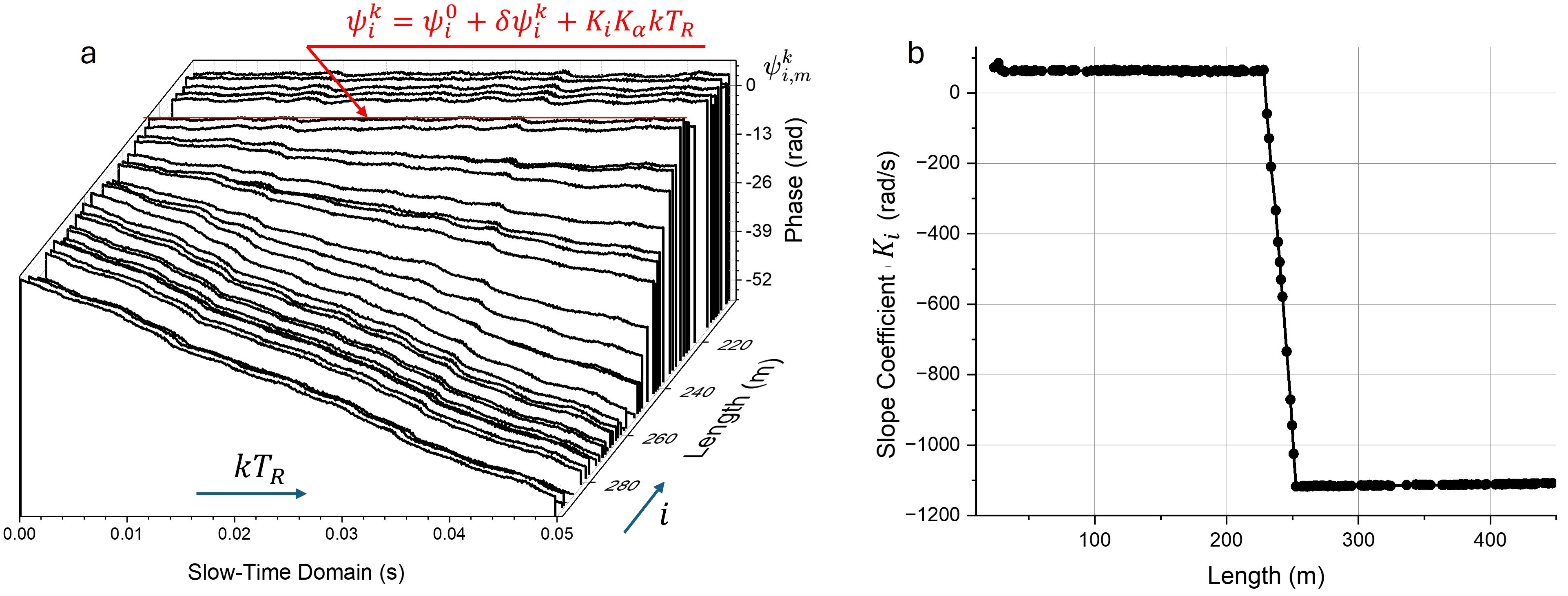}
\caption{a) Phase $\psi^k_{i,m}$ change with Slow-Time Domain $k$ for different positions of the FUT in the vicinity of temperature affected zone; b) Slope coefficients $K_i$ of the Phase $\psi^k_{i,m}$ for different positions of FUT $z^0_i$.}
\label{fig:phase_with_slowtime}
\end{figure}

Fig.~\ref{fig:phase_with_slowtime}(a) illustrates the slow-time evolution of $\psi_{i,m}^k$ in the vicinity of the thermally affected fiber segment. The corresponding coefficients $K_i$ as a function of fiber position $z_i$ are shown in Fig.~\ref{fig:phase_with_slowtime}(b). Since $K_i$ represents the cumulative temperature rate, its spatial distribution directly reflects the extent of the temperature-affected zone. In Fig.~\ref{fig:phase_with_slowtime}(b), this distribution exhibits an approximately rectangular profile, being nonzero only within the temperature-affected segment of the fiber.

\begin{figure}[b!]
\centering
\includegraphics[width=6.4cm]{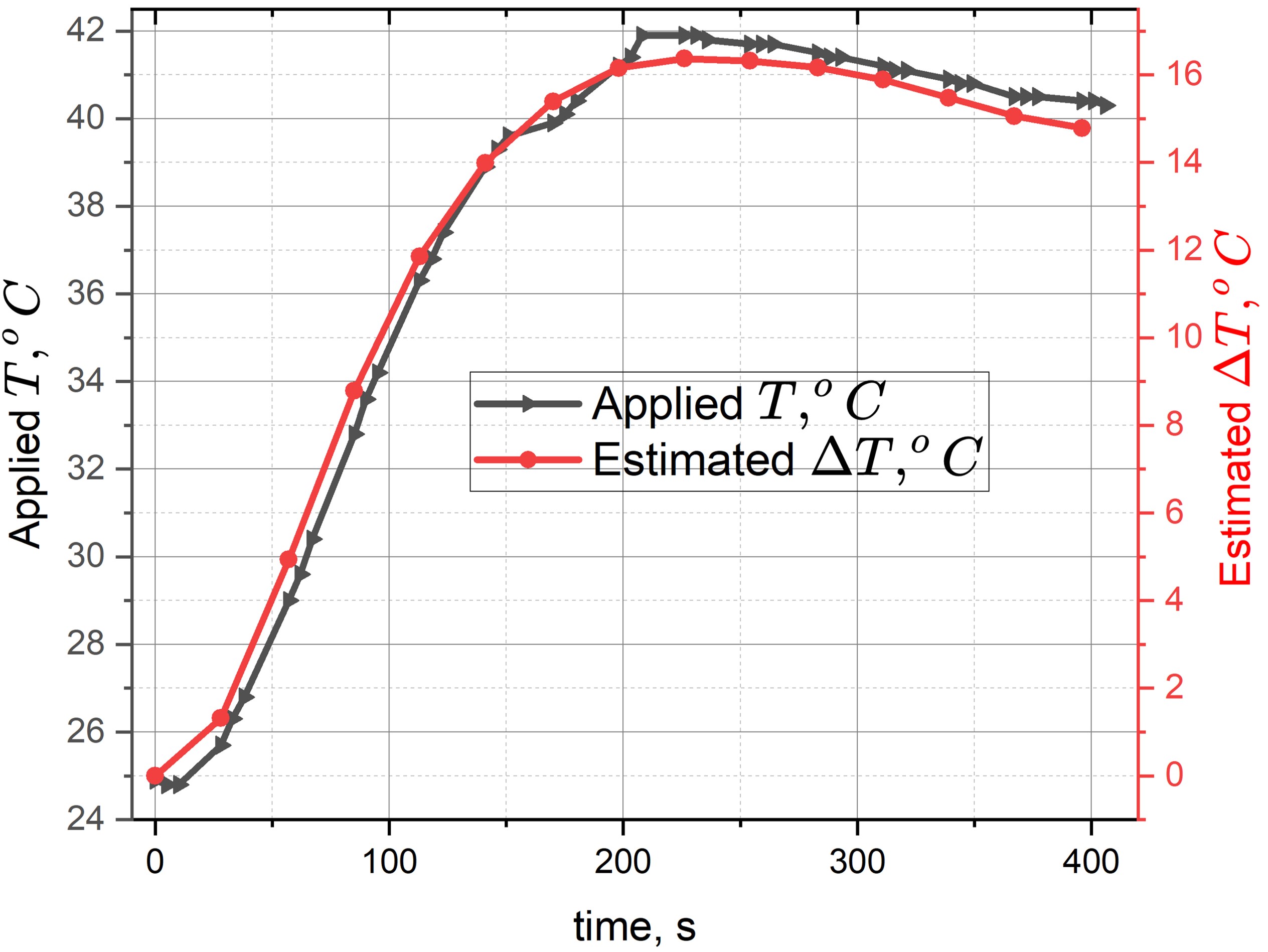}
\hspace{0.3cm}
\includegraphics[width=6.4cm]{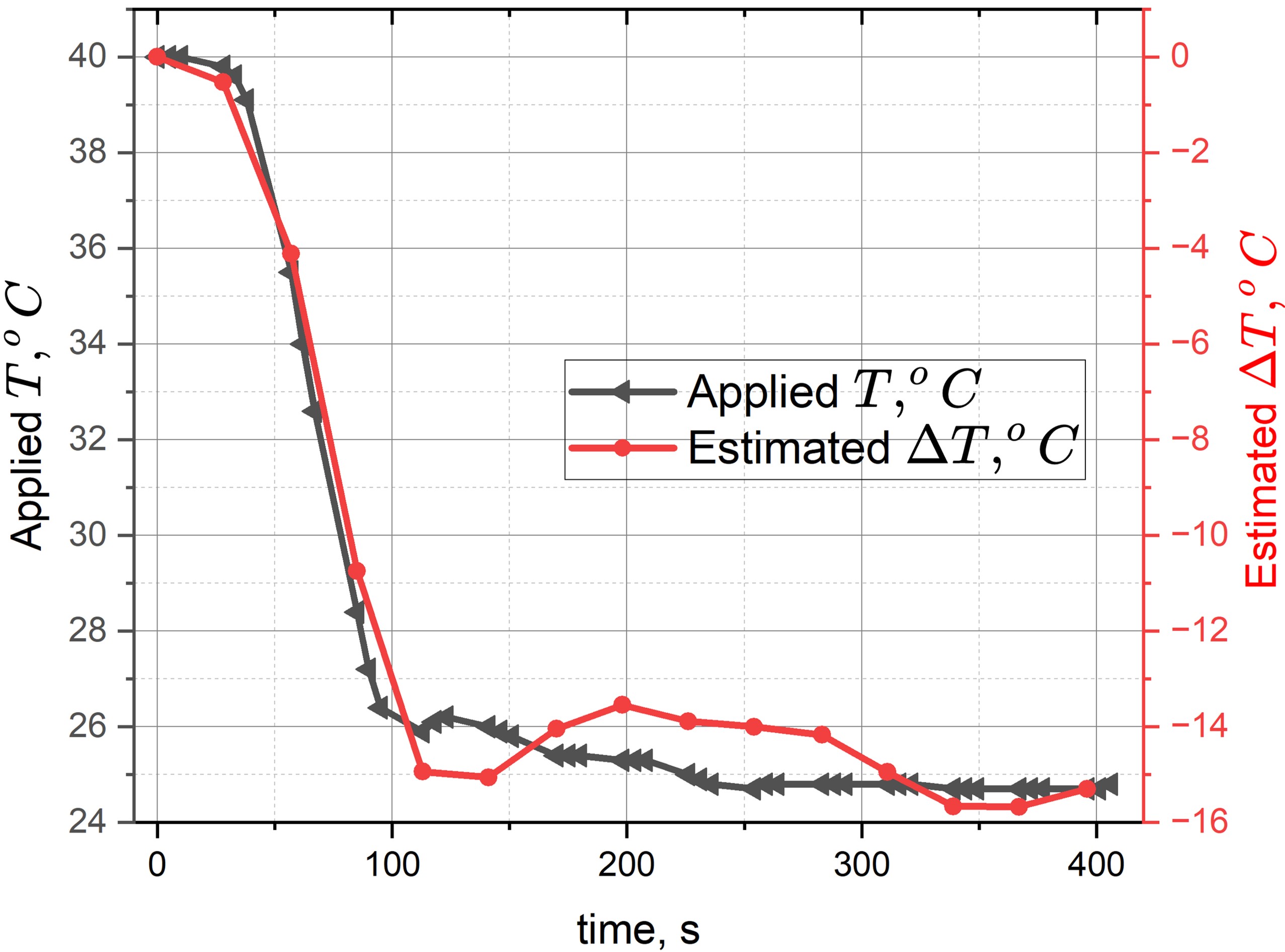}
\caption{Applied versus recovered temperature profiles at a selected segment of the FUT during heating (left) and cooling (right) scenarios.}
\label{fig:temp_recovery}
\end{figure}

For a rectangular temperature-rate profile, the coefficient $K_i$ exhibits a linear dependence on fiber position and can be expressed as
\begin{equation}
\label{eq:coef_linear}
K_i = \left(z_i^0 - z_{\mathrm{start}}\right)\,\dot{T}_i, \,\,\text{for}\,\, z^0_i \in \left[z_{start}, z_{end}\right]
\end{equation}
from which the local temperature rate $\dot{T}_i$ can be directly extracted by linear fitting. Once a sequence of timestamps and their corresponding temperature-rate estimates is obtained for a given segment of the FUT, the temporal evolution of the temperature can be reconstructed.

Accurate temperature reconstruction requires sufficiently frequent sampling, such that the temporal resolution is adequate to capture the dynamics of the temperature variation. Each sample represents an estimate of the temperature rate at a specific time instant, and various numerical methods can be employed to recover the temperature profile from these data.

In this work, we adopt an approach based on the assumption that any three consecutive temperature-rate samples can be locally approximated by a parabolic function. The coefficients of each parabola are determined using the known temperature-rate values at the corresponding timestamps, enabling smooth and robust reconstruction of the temperature evolution.

Fig.~\ref{fig:temp_recovery} illustrates this concept: the true temperature change over time is shown as a black curve, while the reconstructed temperature profile using the proposed method is shown in red. The close agreement between the curves demonstrates the feasibility and accuracy of the temperature recovery approach.

\section{Conclusion}

In this work, we have developed a theoretical framework that describes how the Rayleigh backscattered signal in \(\phi\)-OTDR is modified by locally applied temperature perturbations along the sensing fiber. Within this framework, we show that external temperature changes affect both the phase and the amplitude of the backscattered field simultaneously, whereas polarization-induced fading manifests primarily as a slow variation of the amplitude alone. This distinction can be exploited to discriminate genuine temperature events from polarization effects. We further clarify that the measured phase has an inherently integrative nature, capturing the cumulative effect of temperature variations along the propagation path, while the amplitude responds predominantly to local temperature changes.

Based on this analysis, we derived a practical closed-form expression that quantifies how temperature variations affect the measured phase at a given fiber location. This phase–temperature relation provides the foundation for two signal-processing algorithms: one for detecting and spatially bounding temperature events, and another for reconstructing the corresponding temperature-change profiles in time and space. We validated the proposed framework and algorithms experimentally, demonstrating close agreement between the reconstructed temperature distributions and independent thermometer measurements, with uncertainties below \(1.5^{\circ}\mathrm{C}\). Beyond these immediate results, the derived relations between temperature, Rayleigh amplitude, and phase offer a physically grounded basis for accurate numerical modeling of \(\phi\)-OTDR systems and for designing data-driven approaches, including neural-network-based methods, that exploit this structure for enhanced distributed temperature sensing.

\section{Back matter}


\begin{backmatter}
\bmsection{Funding}
This work was supported by the Horizon Europe Framework Programme
under the SoFiN Project (Grant No.\ 101093015), the ICON Project
(Grant No.\ 101189703), and Villum Fonden (VI-POPCOM VIL54486,
OPTIC-AI VIL29334).



\bmsection{Disclosures}
The authors declare no conflicts of interest.





\bmsection{Data Availability}
Data underlying the results presented in this paper are not publicly available at this time but may be obtained from the authors upon reasonable request.

\bigskip

\noindent Data availability statements are not required for preprint submissions.

\end{backmatter}

\bibliography{sample}






\end{document}